\documentclass[%
 reprint,
superscriptaddress,
 amsmath,amssymb,
prapplied,
floatfix,
longbibliography
]{revtex4-2}

\usepackage{graphicx}
\usepackage{color}
\usepackage{hyperref}
\usepackage{siunitx}
\usepackage{textcomp}
\usepackage{xfrac}
\usepackage{gensymb}
\usepackage{miller}

\begin{document}

\title{Imaging damage in steel using a diamond magnetometer}

\author{L. Q. Zhou}
\affiliation{Department of Physics, University of Warwick, Gibbet Hill Road, Coventry CV4 7AL, United Kingdom}
\author{R. L. Patel}
\affiliation{Department of Physics, University of Warwick, Gibbet Hill Road, Coventry CV4 7AL, United Kingdom}
\affiliation{EPSRC Centre for Doctoral Training in Diamond Science and Technology, University of Warwick, Coventry, CV4 7AL, United Kingdom}
\author{A. C. Frangeskou}
\altaffiliation{Current address: Lightbox Jewelry, Orion House, 5 Upper St. Martin’s Lane, London, WC2H 9EA, United Kingdom}
\affiliation{Department of Physics, University of Warwick, Gibbet Hill Road, Coventry CV4 7AL, United Kingdom}
\author{A. Nikitin}
\altaffiliation{Current address: School of Engineering, University of Warwick, Coventry, CV4 7AL, United Kingdom}
\affiliation{Department of Physics, University of Warwick, Gibbet Hill Road, Coventry CV4 7AL, United Kingdom}
\author{B. L. Green}
\affiliation{Department of Physics, University of Warwick, Gibbet Hill Road, Coventry CV4 7AL, United Kingdom}
\author{B. G. Breeze}
\affiliation{Department of Physics, University of Warwick, Gibbet Hill Road, Coventry CV4 7AL, United Kingdom}
\affiliation{Spectroscopy RTP, University of Warwick, Gibbet Hill Road, Coventry CV4 7AL, United Kingdom}
\author{S. Onoda}
\affiliation{National Institutes for Quantum and Radiological Science and Technology, 1233 Watanuki, Takasaki, Gunma 370-1292, Japan}
\author{J. Isoya}
\affiliation{Research Center for Knowledge Communities, University of Tsukuba, Tsukuba, Ibaraki 305-8550, Japan}
\author{G. W. Morley}
\email{Gavin.Morley@warwick.ac.uk}
\affiliation{Department of Physics, University of Warwick, Gibbet Hill Road, Coventry CV4 7AL, United Kingdom}
\affiliation{EPSRC Centre for Doctoral Training in Diamond Science and Technology, University of Warwick, Coventry, CV4 7AL, United Kingdom}

\begin{abstract}
We demonstrate a simple, robust and contactless method for non-destructive testing of magnetic materials such as steel. This uses a fiber-coupled magnetic sensor based on nitrogen vacancy centers (NVC) in diamond without magnetic shielding. Previous NVC magnetometry has sought a homogeneous bias magnetic field, but in our design we deliberately applied an inhomogeneous magnetic field. As a consequence of our experimental set-up we achieve a high spatial resolution: 1~mm in the plane parallel and 0.1~mm in the plane perpendicular to the surface of the steel. Structural damage in the steel distorts this inhomogeneous magnetic field and by detecting this distortion we reconstruct the damage profile through quantifying the shifts in the NVC Zeeman splitting. This works even when the steel is covered by a non-magnetic material. The lift-off distance of our sensor head from the surface of 316 stainless steel is up to 3~mm.
\end{abstract}

\date{\today}
\maketitle

\section{Introduction}
\label{sec:level1}

The optically-detected magnetic resonance (ODMR) of nitrogen-vacancy centers (NVC) in diamond can be used as a magnetic sensor~\cite{DOHERTY20131,Barry2020}. Key strengths of NVC magnetometry are the high dynamic range~\cite{Clevenson2018}, operation in wide temperature ranges~\cite{PhysRevX.2.031001}, suitability for high radiation environments~\cite{BACHMAIR2016370}, and chemical inertness. The property of ODMR allows nanoscale resolution magnetometry, when employing single centers~\cite{Balasubramanian2008,Maze2008,Taylor2008a}. Conversely, an ensemble of NVC allows for higher sensitivities at the expense of the spatial resolution~\cite{Pham_2011,Rondin2014,Schloss2018,Barry2020} with high sensitivities achieved for both DC~\cite{Clevenson2015,Barry2016a,Schloss2018,Webb2019} and AC frequencies~\cite{Wolf2015,LeSage2012}. A range of applications have been demonstrated from single neuron action potential detection~\cite{Barry2016a} to eddy-current-induced magnetic field detection of conductive samples~\cite{PhysRevApplied.11.014060} for material analysis.

The identification of structural defects through the use of magnetic flux leakage (MFL) is a  non-destructive testing technique that is among the most-used methods for yielding information about the nature of unknown defects in magnetic materials~\cite{doi:10.1177/146442070121500403,WANG2012382,GHOLIZADEH201650,doi:10.1243/09544097JRRT209,DWIVEDI20183690,JILES1988311,JILES199083,Mandache_2003,AFZAL2002449}. MFL measurements involve magnetically saturating the target material. If there is no damage the magnetic flux lines are unperturbed, however, if there is a flaw magnetic flux will leak out of the material. MFL measurements have found applications in industries where the corrosion of magnetic material, such as steel, will eventually lead to significant material loss and in particular have been heavily utilized by the oil and gas industries in pipeline inspection gauges to minimize the need for costly excavations~\cite{Ryu_2002,Shi2015,MFL_2007,ZUOYING200661,5274410,1704562,6606345,6599681,7752913}. Several different magnetic field sensors are employed in industry with the key technologies being induction coils, Hall probe sensors and magnetic flux gates with each offering distinct advantages and disadvantages~\cite{Shi2015}. Hall sensors in particular have found great usage for MFL measurements due to their low cost~\cite{Shi2015,LIANG2017127,HANED2003216}, however these sensors suffer from voltage drift, even in the absence of a magnetic field and thus require compensation and offer limited sensitivity in comparison to other sensors~\cite{Shi2015,LIANG2017127}. In addition, to effectively utilize the techniques of MFL the magnetic material under inspection must undergo magnetic saturation~\cite{Shi2015,Mandache_2003,AFZAL2002449}. This requirement sometimes makes it difficult to utilize the technique in the field.

Two quantum systems have been used recently for imaging structural damage in metals~\cite{PhysRevApplied.11.014060,Bevington2018}. The first used an atomic vapor cell to achieve a spatial resolution of 0.1~mm~\cite{Bevington2018}. However, this required the use of a sensitive commercial fluxgate magnetometer and 1 m electromagnet coils to null the background magnetic field. Chatzidrosos {\em et al.}~\cite{PhysRevApplied.11.014060} demonstrated an NVC magnetometer design without microwaves~(MW) to remove the problem of the ODMR microwaves interfering with the conductive materials under study. This required a relatively high external bias magnetic field of 102.4~mT supplied by a large water-cooled electromagnet. 

Our sensor is based on ODMR of NVC but using MW excitation means that we only need a low applied bias magnetic field from two permanent magnets. Our sensor head design prevents MW leakage with a small Faraday shield and does not use any compensation coils. Under our configuration we detect magnetic flux profile perturbations that arise in the permanent magnetic bias fields that are used to induce Zeeman splitting of the NVC; a simulated 2D magnetic flux profile of the setup is available in the supplemental information (S.I.). These perturbations occur when the magnetic properties of the material are changed, such as by structural defects due to corrosion. This novel method of flux detection allows for reconstruction of the profile of defects in magnetic materials such as common steels. Furthermore, as the techniques employed do not require magnetic saturation we thus avoid some of the limitations inherent to magnetic flux leakage measurements and hence a distinct advantage is offered compared to standard magnetic flux leakage measurements. We are still able to provide an accurate reconstruction of the examined defects in an unshielded environment. Our fiber-coupled design with a small sensor head increases the flexibility for practical applications.  The large lift-off of up to 3~mm allows this sensor to examine magnetic materials directly even when coated with a thick non-magnetic layer.

\section{Experimental Details}
\begin{figure}[t]
    \centering
    \includegraphics[width=1\linewidth]{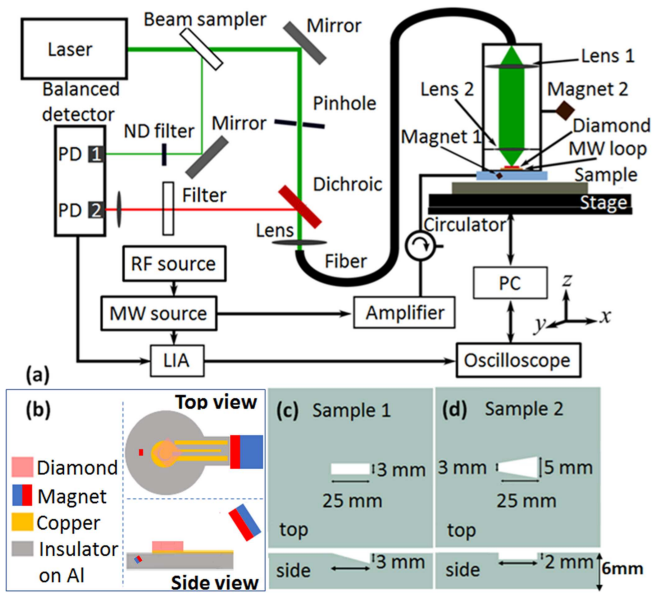}
    \caption{(a) A schematic of the experimental set-up, including the neutral density (ND) filter, microwave (MW) and radio frequency (RF) sources, lock-in amplifier (LIA) the two photodiodes (PD's) of the balanced detector. (b) Schematic of the microwave antenna used for MW delivery. A pair of magnets were used to induce Zeeman splitting (the blue and red colors are used to illustrate the differing magnetic poles). (c) and (d) are the 316 stainless steel plates used in this work. Plate dimensions were 150~mm by 150~mm, here only a small area where the defects were introduced onto the plates are shown for clarity; the size of schematics are not to scale.}
    \label{Experiment_samples}
\end{figure}

We have previously achieved a sensitivity of 310~$\pm$~20~pT/$\sqrt{\textrm{Hz}}$ in the frequency range of 10 to 150~Hz using an isotopically-purified $^{12}$C diamond with a similar configuration~\cite{Patel2020}. However, this configuration had poor spatial resolution for imaging. Here we solve this problem by introducing a 1~mm permanent magnet 2~mm from the diamond. While improving the spatial resolution (due to the steel experiencing an inhomogeneous bias field), the resulting inhomogeneous bias field on the diamond impairs the magnetic sensitivity. This means there is no need to use $^{12}$C diamond because our measurements are not limited by the natural abundance of $^{13}$C impurities. However, improving the sensitivity is still useful as it allows a higher lift-off from the steel. 

The experimental set-up is shown in figures~\ref{Experiment_samples}(a) and \ref{Experiment_samples}(b), see S.I. for more details about the experimental setup and diamond information. Figures~\ref{Experiment_samples}(c) and ~\ref{Experiment_samples}(d) show the two test samples used; each is made of 316 stainless steel with different slot defects. Sample 1 contains a slot with a 3~mm width with a gradient depth from 0 to 3~mm. Conversely, sample~2 has a trapezoidal shape with a width ranging from 3 to 5~mm, and a fixed depth of 3~mm. Both samples were mounted onto the bed of a scanning stage whilst the sensor head was affixed to the $z$-axis component stage. The scanning stage was used to enable two-dimensional scanning in the $x$ and $y$ axes with a different lift-off distance in the $z$-axis. All data from the balanced output were digitized by a Zurich MFLI DC-500~kHz lock-in amplifier (LIA), with the MW frequency modulated. 

\section{Results and Discussion}

\begin{figure}
    \centering
    \includegraphics[width=1\linewidth]{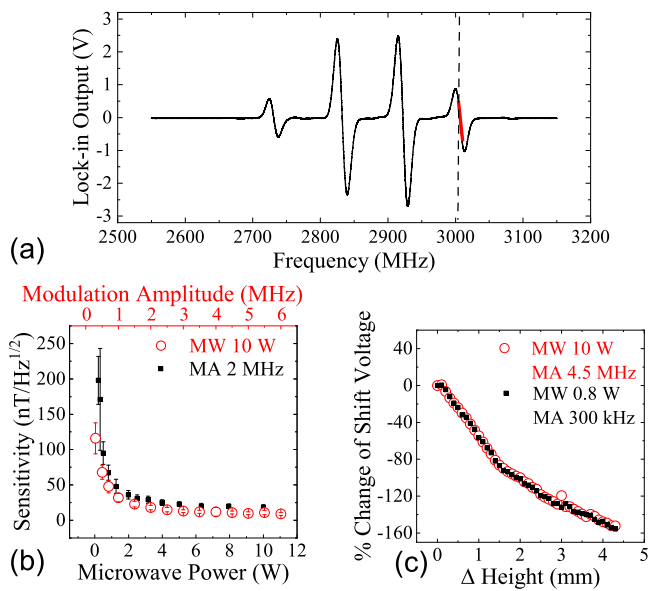}
    \caption{(a) Optically-detected magnetic resonance (ODMR) signal from the lock-in amplifier (LIA) output, (b) dependence of sensitivity on input microwave (MW) power (black square) and modulation amplitude (MA) (red circle). (c) Changes to the LIA voltage output whilst monitoring a fixed MW frequency when the lift-off distance between the sensor head and a damage free 316 stainless steel sample is changed is shown. This was taken under two differing sensitivity settings. The lift-off was 0.2 mm for $\Delta$Height = 0~mm.}
    \label{ODMR_Sensitivity_liftoff}
\end{figure}

The sensitivity of NVC magnetometers is highly dependent on the orientation of the magnetic fields relative to the NVC symmetry axis and the MW delivery parameters, such as the MW power, frequency modulation amplitude and modulation frequency~\cite{Patel2020}. An NVC ODMR spectrum where the bias magnetic field is aligned along a $\hkl<$111$>$ orientation is shown, see figure~\ref{ODMR_Sensitivity_liftoff}(a). A MW power of 10~W was used with a frequency modulation amplitude of 4.5~MHz and a modulation frequency of 3.0307~kHz; these parameters were used for all measurements relating to structural defect quantification of the stainless steel samples. The highlighted region at $\sim$3~GHz is the region of the ODMR feature where all scanning measurements of the 316 stainless steel plates were performed. The optimum parameters of operation, see figure~\ref{ODMR_Sensitivity_liftoff}(b), were found through variation of the MW power, between 0.2~W and 10~W at a fixed frequency modulation amplitude and modulation frequency. Conversely, the optimum frequency modulation amplitude was found through variation between 300~kHz and 6~MHz whilst the MW power and modulation frequency were fixed. For each parameter an ODMR spectrum was taken and linear fits applied around the central frequency of the ODMR feature in conjunction with a one second fast Fourier transform (FFT) of the voltage output of the central frequency. The resulting sensitivity is shown in figure~\ref{ODMR_Sensitivity_liftoff}(b), where each sensitivity is the mean of 96 FFTs; the errors are the standard deviation.

The resonance of the NVC shifts when the value of the external magnetic field changes due to Zeeman shifts of the $m_{s}$~=~$\pm1$ energy levels. Changes to the NVC resonance causes a change in the fluorescence and this correspondingly changes the voltage output of the LIA. In figure~\ref{ODMR_Sensitivity_liftoff}(c) changes to the  
external magnetic field incident upon the NVC were caused by changes introduced to the distance between the sensor head and the surface of the 316 stainless steel plates in a damage-free area. The distance of the sensor head from the surface of the ferromagnetic sample was increased and the voltage changes to the LIA output were found (further details concerning the NVC ODMR changes are discussed in the S.I.), the reference point used is indicated by the dashed line in figure~\ref{ODMR_Sensitivity_liftoff}(a), this corresponds to an LIA output value of 0.6~V. Although we could use any value as the reference voltage provided it is along the linear region of the NVC ODMR resonance, and it is typical to use the zero crossing point, we chose to use 0.6~V as the starting reference voltage to allow for a higher lift-off distance in our measurements. The calibration in figure~\ref{ODMR_Sensitivity_liftoff}(c) was performed upon a blank area of the 316 stainless steel sample. In this instance the reference frequency was chosen as the lowest value of the MW frequency of the highlighted feature in figure~\ref{ODMR_Sensitivity_liftoff}(a). It is evident that the relationship between the Zeeman-induced change upon the defects is not linear with respect to the distance. This is attributed to the dipolar field pattern from the permanent magnets used~\cite{Petruska2013}. The trend of changes to the LIA output during the lift-off process are independent of the sensitivity, see  figure~\ref{ODMR_Sensitivity_liftoff}(c); the trends are near identical regardless of the MW parameters used. This is expected as provided measurements are in the linear region of the ODMR feature, the differences should be entirely due to the bias magnets and their position relative to the active area of sensing within the diamond. As the differences caused are directly due to magnetic flux distortions it is entirely possible to use this technique to map out the surface structure of magnetic materials.

\begin{figure}
    \centering
    \includegraphics[width=1\linewidth]{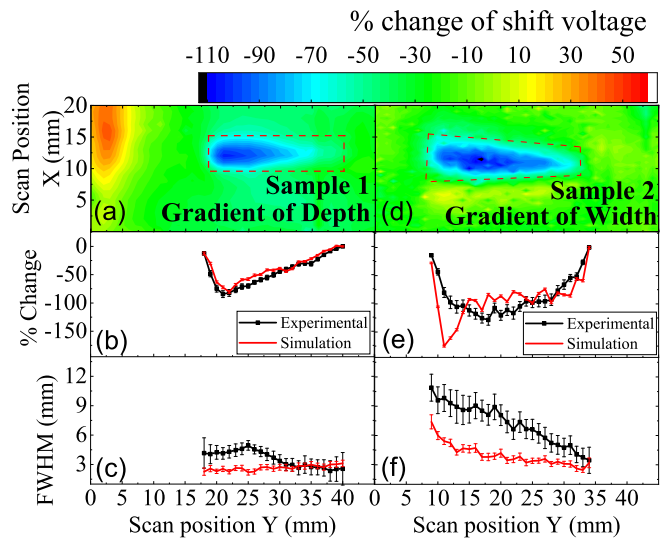}
    \caption{2D-scan results for both steel samples 1 and 2 are shown in (a) and (d) respectively. The scan dimensions were 45~mm~$\times$~20~mm. Lorentzian fits to the cross sectional profile of (a) and (d) were performed and are shown in (b) and (e) respectively. The full width at half maximum (FWHM) of the cross section of both defects imaged by the sensor, (c) and (f) are shown for sample~1 and sample~2 respectively. COMSOL simulations were performed and the results are shown in 
    (b), (c), (e) and (f).}
    \label{2Dscan_results_1}
\end{figure}

The voltage change response of the magnetic field sensor caused by the steel samples surface structure are shown in figure~\ref{2Dscan_results_1}. The scan resolution in the $x$ and $y$ axes were 1~mm. To prevent interference from the motors of the scanning stage a dwell time of 1~s was implemented before the data acquisition process began. This led to a total scanning time of 2-3 seconds per point leading to a total scan time of approximately 66 minutes for a 20~mm by 45~mm scan. The total scan time for the maps in figure~\ref{2Dscan_results_1} were 66 minutes. Both defects are mapped out and clearly visible as shown in figure~\ref{2Dscan_results_1}(a) and ~\ref{2Dscan_results_1}(d). The lift-off distance between the surface of the sample and the sensor head from the base of the aluminum PCB antenna was 0.2~mm for all scans in both figure~\ref{2Dscan_results_1}(a) and \ref{2Dscan_results_1}(d). Using the voltage difference and the calibration performed in figure~\ref{ODMR_Sensitivity_liftoff}(c) it is possible to evaluate differences in the depth of the features. We believe that the increase in normalized voltage output in figure~\ref{2Dscan_results_1}(a) in the top left corner is due to the larger magnet being close to the sample edge.

Figure~\ref{2Dscan_results_1}(a) and figure~\ref{2Dscan_results_1}(d) reveal differences in profile and show that a ferromagnetic sample can be differentiated based on its depth and width. 
Though the differences in the 2D profiles in figure~\ref{2Dscan_results_1}(a) and figure~\ref{2Dscan_results_1}(d) can identify the differing nature of the samples, to further highlight these differences and confirm the distinction between depth and width, the differences in the profile of the cross sections from figures~\ref{2Dscan_results_1}(a) and ~\ref{2Dscan_results_1}(d) were analyzed by performing Lorentzian fits across the width spanned by the defect. It is expected for a sample where the depth changes but not the width that the magnitude of change to the LIA voltage output will increase across the length of the defect but this change will not be reflected in changes to the full width at half maximum (FWHM) which should remain fairly constant. This is demonstrated by figure~\ref{2Dscan_results_1}(c) where the FWHM is for the most part constant while figure~\ref{2Dscan_results_1}(f) shows that the amplitude increases in an almost linear fashion with increases with the depth of the defect. The rapid change in the shift voltage in figure~\ref{2Dscan_results_1}(b) is due to the boundary of the defect edge where the depth change is maximum. In contrast, it is expected that the magnitude of change for a sample whose depth is unchanged will be constant across the length of the defect while the FWHM will change with a larger width resulting in a larger FWHM value. This is confirmed by figure~\ref{2Dscan_results_1}(f) which shows a constant increase in the FWHM as the width of the defect increases but relatively constant voltage changes as shown in figure~\ref{2Dscan_results_1}(e). COMSOL simulations of the experiment (see S.I. for details) yield defect cross-sections and magnitude signals which are in overall agreement with the experiment. These results are shown in figures~\ref{2Dscan_results_1}(b), (c), (e), (f). The steel magnetic permeability used was 1.02 in the simulation as discussed in the S.I. 
\begin{figure}[t]
    \centering
    \includegraphics[width=1\linewidth]{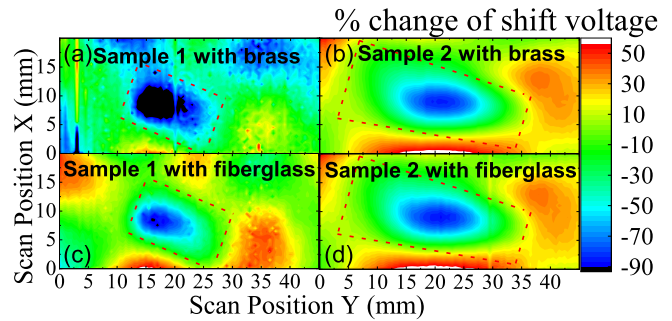}
    \caption{Two dimensional (2D) scans of samples 1 and 2 when covered with nonmagnetic material covering. The scan dimensions were 45~$\times$20~mm. The structural defects present in samples 1 and 2 are encapsulated with the dashed line. (a) and (c) are from sample 1 while (b) and (d) are from sample 2. For (a) and (b) the steel is covered with 1.5 mm of brass while for (c) and (d) the steel is covered with 2 mm of fiberglass.}
    \label{cross_section_analysis}
\end{figure}

Steel corrosion under insulation is an important global problem and we demonstrate it is possible to map the defects even when they are covered by non-magnetic materials as shown in figures~\ref{cross_section_analysis}(a)-\ref{cross_section_analysis}(d). The lift-off distance was 3~mm for all scans in figure~\ref{cross_section_analysis} - when compared to those in figure~\ref{2Dscan_results_1}, the larger lift-off means that the magnetic flux from the smaller magnet is incident over a larger area of the steel. This leads to a worse spatial resolution in the plane parallel to the sample surface: the $x$ and $y$ directions. There are also more spurious signals than in figure~\ref{2Dscan_results_1} which shall be addressed in future research. In the $z$-axis we achieve a 0.1~mm spatial resolution while for the $x$-axis and $y$-axis we reach $\sim$1~mm which we believe is limited by the 1~mm size of the small magnet. The minimum lift-off in the $z$-axis was set by the nominal resolution of the scanning stage in the $z$-axis; increments lower than 0.1~mm were not possible. It may be possible to enhance the spatial resolution either through using a smaller magnet, or one with that is shaped to have a sharp point facing the steel.

\section{Conclusion}
In this work, we demonstrate a novel method of imaging defects in magnetic materials using a compact sensor based on an ensemble of nitrogen-vacancy color centers in diamond. This device can be used to detect structural defects in magnetic materials and aid in their quantification even when covered with non-magnetic materials. This is particularly useful as corrosion under insulation is an important global problem. Reducing the size of the 1~mm cube bias magnet may improve the spatial resolution of our measurements. Furthermore, as the sensor head is based on diamond, with further improvements to the design, we expect that it would be suitable for operation in radioactive environments~\cite{BACHMAIR2016370}, and  operation up to 300~{\textdegree}C~\cite{PhysRevX.2.031001} is believed to be possible.

\begin{acknowledgments}
The authors thank Luke Johnson for materials processing, Jeanette Chattaway, Matty Mills and Lance Fawcett of the Warwick Physics mechanical workshop, and Robert Day and David Greenshield of the Warwick Physics electronics workshop. The authors are particularly grateful for insightful discussions with Steve Dixon, Jozef Tzock, Zhichao Li and Rachel Edwards (University of Warwick) concerning non-destructive testing. R. L. P's PhD studentship is funded by the EPSRC Centre for Doctoral Training in Diamond Science and Technology (Grant No. EP/L015315/1). L. Q. Z. is supported by an EPSRC Impact Acceleration Account grant from Warwick Ventures. The use of the Spectroscopy Research Technology Platform facilities at the University of Warwick is acknowledged. B. L. G is supported by the Royal Academy of Engineering. This work was supported by the EPSRC Quantum Technology Hub NQIT (Networked Quantum Information Technologies - Grant No. EP/M013243/1) and QCS (Quantum Computing and Simulation - Grant No. EP/T001062/1). G. W. M. is supported by the Royal Society.
\end{acknowledgments}

\bibliography{library}

\end{document}